# Improving Suppression to Reduce Disclosure Risk and Enhance Data Utility

**Marmar Orooji and Gerald M. Knapp**
**Louisiana State University**
**Baton Rouge, LA 70803, USA**

## Abstract

In Privacy Preserving Data Publishing, various privacy models have been developed for employing anonymization operations on sensitive individual level datasets, in order to publish the data for public access while preserving the privacy of individuals in the dataset. However, there is always a trade-off between preserving privacy and data utility; the more changes we make on the confidential dataset to reduce disclosure risk, the more information the data loses and the less data utility it preserves. The optimum privacy technique is the one that results in a dataset with minimum disclosure risk and maximum data utility. In this paper, we propose an improved suppression method, which reduces the disclosure risk and enhances the data utility by targeting the highest risk records and keeping other records intact. We have shown the effectiveness of our approach through an experiment on a real-world confidential dataset.

## 1. Introduction

Individual-level data (aka microdata), which can be used for detailed modeling and machine learning, usually contains private and sensitive information about individuals and thus generally cannot be made available for public access, at least no without processing to provide privacy guarantees. Attributes existing in microdata are generally classified into three categories: direct identifiers, quasi-identifiers (QID), and sensitive attributes (SA). Direct identifiers are unique to a person such as social security number. QIDs are not unique per person but when considered as a set, the combination of QID field values can be used to identify individuals with high probability. Examples are birthdate, gender, and zip code. SAs contain private information about individuals such as healthcare information. An "adversary" can disclose SA values about a person, called the "victim", even when direct identifiers are removed if they know the QID values of the victim and can match them against QIDs appearing in the sensitive microdata. This raises the question of how data owners can share their data for research purposes while not violating individuals' privacy. This problem has been recently studied in depth in the Privacy Preserving Data Publishing (PPDP) area, in which microdata is anonymized such that the privacy of individuals in the confidential dataset is preserved up to a certain level.

In addition to privacy preservation, it is of crucial not to lose too much information due to the anonymization operations because data users are always seeking to have as much information as possible. The challenge is the trade-off between preserving privacy and data utility, i.e., the more we anonymize the data to better preserve individual's privacy, the more information the data loses and the less data utility it preserves. This problem has been the subject of recent studies (discuss below) to improve privacy techniques to achieve higher data utility while not losing privacy.

In this paper, we present a novel approach for reducing disclosure risk and enhancing data utility, which is effective on datasets with large number of SAs. We improve the anonymization operations by targeting only records at high risk of disclosure and keeping other records intact. We first overview a novel disclosure risk metric, which measures disclosure risk of each record. We then identify records with risk value greater than a threshold as high-risk records, and only anonymize these records to reduce disclosure risk.

## 2. Literature Review

Various privacy models have been proposed in the PPDP area, each specifies a privacy requirement that is achieved by employing anonymization operations on the sensitive microdata. "k-anonymity" is a fundamental, widely used privacy model which requires anonymizing dataset such that groups of at least k records with equal QID value set are created [1]. This produces a dataset where an adversary finds at least k matched records for a victim for any QID value set, and thus the probability of uniquely identifying an individual's record decreases to $1/k$. Anonymization operations employed for this model were generalization and suppression. Generalization replaces the original values with more generalized ones. For example, the age of a person can be generalized to 10 years intervals, so a person who is 32 years old might be transformed to the age value [30-40]. Suppression takes place at record or value level, meaning it deletes either the whole record or some values of a record. "$\ell$-Diversity" [2] and "t-closeness" [3] are well-known privacy models which add other requirements to k-anonymity to better thwart disclosure attacks. $\ell$-Diversity restricts each group of equal QID values to have diverse SA values and t-closeness requires the distribution of SA values within each group to be close to the distribution of SA values within the entire dataset.

Early privacy models considered only a single sensitive attribute at a time, and extending these models to preserve multiple sensitive attributes incurs substantial information loss. In recent studies, researchers have focused on handling multiple SAs while preserving data utility, which is practical for real-world sensitive datasets. For instance, Wang and Zhu presented a novel algorithm that can thwart attacks to two SAs at a time [4]. Li et al. proposed a new approach called "slicing" to improve data utility of privacy techniques [5], and Susan and Christopher used this approach to address several SAs [6]. Some works have limited their approach to handle multiple SAs of the same type, e.g., multiple numeric SAs [7] or categorical ones [8].

In order to measure data utility of an anonymized dataset, various data utility metrics have been proposed in the literature. In this work, we use Normalized Certainty Penalty (NCP) [9] as a measure to evaluate data utility of the anonymized dataset obtained from our proposed approach. NCP measures information loss by penalizing attribute values, which are transformed to more generalized values after anonymization. Some surveys have reviewed different privacy models, anonymization operations, privacy techniques, and data utility metrics [10, 11].

## 3. Methodology

Our approach improves the process of anonymizing sensitive microdata in terms of reducing disclosure risk and enhancing data utility. The novel idea is to present a disclosure risk metric that measures disclosure risk of each record of microdata, and identifying records at high disclosure risk. Then, choosing suppression as the anonymization operation, we propose a value suppression algorithm, which deletes values with the highest contribution in the disclosure risk measure for each high-risk record. Thus, our approach decreases disclosure risk since it suppresses high risk factors, and increases data utility because it only anonymizes high-risk records and keeps other records intact.

### 3.1 Disclosure Risk Measure

Our proposed disclosure risk measure is at record level and defined based on risk assessment, which is likelihood times consequence. Likelihood of a record $r$ indicates the likelihood of being identified by an adversary and consequence specifies the sensitivity of the private information of $r$ being revealed, if $r$ is identified. For calculating likelihood and consequence, we split attributes into two sets; known set of attributes vs. unknown set of attributes. The known set contains attributes, which an adversary knows about a victim and uses to find matched records in microdata. The remaining attributes form the unknown set and contain private information which adversary wants to disclose for a victim. Likelihood and consequence in the proposed risk measure are calculated based on the known set and unknown set, respectively. In order to take into account an adversary with any known and unknown attributes, we are considering all subsets of attributes as possible known sets and their compliment subsets as unknown sets. Since each attribute has 2 possibilities − i.e., being in known set or in unknown set, having $m$ number of attributes, total number of known sets is equal to total number of unknown sets and is equal to $2^m$. Considering all possible known/unknown sets, our proposed disclosure risk measure is defined as:

$$D(r) = \sum_{i=1}^{2^m} L_{KS_i}(r) \times \alpha\, C_{UKS_i}(r) \quad (1)$$

where $KS_i$ is the $i^{th}$ known set, $UKS_i$ is the $i^{th}$ unknown set, and $\alpha$ is consequence coefficient which is a constant value larger than 1 to magnify the effect of consequence in risk value. If a dataset contains large number of attributes, calculating our proposed disclosure risk will be computationally expensive. A heuristic pruning algorithm can be employed to remove known sets which are incurring very low likelihood values [12].

$L_{KS_i}(r)$ indicates the likelihood that an adversary, who knows the attributes in the $i^{th}$ known set for a victim, finds the record $r$ as the matched record for the victim. This consists of two terms − i.e., the probability that $i^{th}$ known set is publicly known, and the uniqueness of $i^{th}$ known set's attribute values for record $r$ in microdata. It is formulated as:

$$L_{KS_i}(r) = \prod_{A_j \in KS_i} PK(A_j) \times \frac{1}{F(r(KS_i))} \tag{2}$$

where, $A_j$ is the $j^{th}$ attribute in $KS_i$, $PK(A_j)$ is the probability of $A_j$ to be publicly known, $r(KS_i)$ is $KS_i$'s attribute values for record $r$, and $F(r(KS_i))$ is the frequency of $r(KS_i)$ in microdata.

The first term in likelihood implies how probable the set of attributes in the $i^{th}$ known set is to be publicly known so that an adversary knows them about a victim. These probabilities are assigned by data publisher intuitively. The second term is the inverse frequency of the $i^{th}$ known set values of the record $r$. This term implies the uniqueness of record $r$ in the whole dataset and has the inverse correlation with the likelihood. If the values are frequent in microdata then the person whom record $r$ belongs to is less likely to be identified by those attributes. Including this term, likelihood is no longer a probability function in our measure. However, it is still a value between 0 and 1 for each record.

$C_{UKS_i}(r)$ denotes how much private information of record $r$, embedded in the $i^{th}$ unknown set, will be revealed, if that record is identified. Data publisher needs to assign sensitivity weights to both attributes and their values to specify how much the information is sensitive and private to be disclosed about a person. This term is derives as:

$$C_{UKS_i}(t) = \sum_{A_j \in UKS_i} W(A_j) \times W\left(r(A_j)\right) \tag{3}$$

where, $r(A_j)$ is the value of $A_j$ for record $r$, $W(A_j)$ is the sensitivity weight of $A_j$, and $\mathrm{W}\left(\mathrm{r}(\mathrm{A_j})\right)$ is the sensitivity weight of $r(A_j)$. Sensitivity weights are constant values between 0 and 1.

### 3.2 Suppression Based on Records' Disclosure Risk Values

We are proposing a value suppression algorithm, which targets records with high disclosure risk value. Using the disclosure risk metric described above, we first calculate the risk of each record and find records with the disclosure risk value more than a threshold $\delta$. Then, for each of these records, we look through the values of $(L_{KS_i} \times \alpha\, C_{UKS_i})$ for all $i$, and find the split of attributes which results in the maximum of $(L_{KS_i} \times \alpha\, C_{UKS_i})$. Having the known set with the most contribution in the risk value, we suppress the values of its attributes for the selected high-risk record.

<u>Value Suppression Algorithm</u>
1. $HR_r = \{r | D(r) > \delta\}$
2. for each $r$ in $HR_r$:
3. $\quad HR_KS_i / HR_UKS_i = \underset{KS_i/UKS_i}{\operatorname{argmax}} \left(L_{KS_i}(r) \times \alpha\, C_{UKS_i}(r)\right)$
4. $\quad$ for each $A_j$ in $HR_KS_i$:
5. $\quad\quad$ Suppress $t(A_j)$

Figure 1. Algorithm for proposed suppression method

The proposed suppression algorithm is shown in Figure 1. High risk records are stored in $HR_r$, and $HR_KS_i/HR\text{-}_UKS_i$ indicates the split of attributes which has the highest contribution in risk value. $HR_KS_i$ shows the known set of the split and $HR_UKS_i$ is the unknown set of the split.

### 3.3 Data Utility Measure on Suppressed Microdata

We use Normalized Certainty Penalty (NCP) metric to measure information loss on the suppressed microdata obtained from our value suppression algorithm. NCP of the anonymized dataset $D'$ with n number of records is derived as:

$$NCP(D') = \frac{1}{n}\sum_{i=1}^{n}\sum_{j=1}^{q} w_j \, NCP_{A_j}(r_i) \tag{4}$$

In (4), $q$ is the number of QIDs. $w_j$ indicates how much information $A_j$ carries for utility and are assigned such that $\sum_{j=1}^{q} w_j = 1$. $NCP_{A_j}(r_i)$ is the assigned penalty for $r_i(A_j)$ value. Since we have used suppression, this value is either an original value or a suppressed one. In the first case, the penalty is 0 and for the latter it is assigned to 1. If we assign equal weight of $1/q$ to each QID, and since $NCP_{A_j}(r_i)$ is either 0 or 1, $NCP(D')$ is simplified as:

$$NCP(D') = \frac{1}{n} \times \frac{1}{q} \times S \tag{5}$$

where, $S$ is the total number of suppressed values in the anonymized dataset.

# 4. Experiment

We have evaluated our approach on a sensitive microdata provided by the Social Research and Evaluation Center at Louisiana State University. The microdata contains 1,009,993 number of students (records) enrolled in Louisiana public schools between 1999-2011 school years, and 27 number of attributes from Louisiana Department of Education (LADOE), Office of Juvenile Justice (LAOJJ), and Department of Corrections (LADOC). This dataset is of high importance to social work researchers [13]. Examples of attributes are age, gender, ethnicity, homelessness, dropout flag, LADOC contact, and LAOJJ contact.

## 4.1 Measuring Disclosure Risk and Applying Value Suppression

The first step is to calculate disclosure risk for each record in the microdata based on our proposed disclosure risk measure. We need to assign each attribute the probability of being publicly known and sensitivity weight, plus sensitivity weight for each value within an attribute. Examples of these values for sample attributes are shown in Table 1. The assigned sensitivity weights are either 0 or 1, and among 27 attributes, we set 1 for 19 attributes and this implies having 19 SAs. In this experiment, we set the consequence coefficient (α) to 100.

Table 1. Examples of assigned publicly known probabilities and sensitivity weights

| Attribute | Publicly Known Probability | Attribute Sensitivity Weight | Value Sensitivity Weight | |
|---|---|---|---|---|
| | | | **Value** | **Weight** |
| Age | 0.05 | 0 | --------- | |
| Gender | 0.8 | 0 | --------- | |
| Dropout Flag | 0.005 | 1 | Yes | 1 |
| | | | No | 0 |
| LADOC Contact Flag | 0.001 | 1 | Yes | 1 |
| | | | No | 0 |

For the parameters specified, we have calculated the disclosure risk measure for each record in the sensitive microdata. The histogram of risk values among the 1,009,993 records is shown in Figure 2 with blue-colored bars. In this experiment, we set 0.01 for the threshold $\delta$ and as shown in Figure 2, about 1.5% of records are at high risk of disclosure. We employed our value suppression algorithm on the high-risk records and obtained an anonymized dataset.

## 4.2 Disclosure Risk Evaluation on the Anonymized dataset

In order to evaluate the reduction in disclosure risk after anonymization, we have calculated the disclosure risk on the anonymized microdata and compared the risk values with those of the original microdata.

Table 2. Number of high-risk records before and after anonymization

| | **Number of high-risk records (risk>0.01)** |
|---|---|
| Original Dataset | 15651 records (1.55%) |
| Anonymized Dataset | 8597 records (0.85%) |

Table 2 shows that the number of high-risk records is reduced by 45%. The comparison of the histogram of risk values among records between the original microdata and anonymized microdata is illustrated in Figure 2. The comparison implies that the number of records having disclosure risk values of more than 50 decreases significantly after value suppression. Therefore, it is concluded that our proposed value suppression anonymization operation is highly effective on preventing occurrences of high disclosure risk values. However, it is shown that the number of records having disclosure risk values of less than 10 and more than 0.5 increases after value suppression. This is justified because of our choice of anonymization operation, which is suppression. When we delete some values of an attribute, the number of occurrences of that value for that attribute becomes smaller. Therefore, the second component of likelihood, which has the inverse correlation with the frequency of attribute, becomes larger for records having their attribute values suppressed and this may increase risk value for some records.

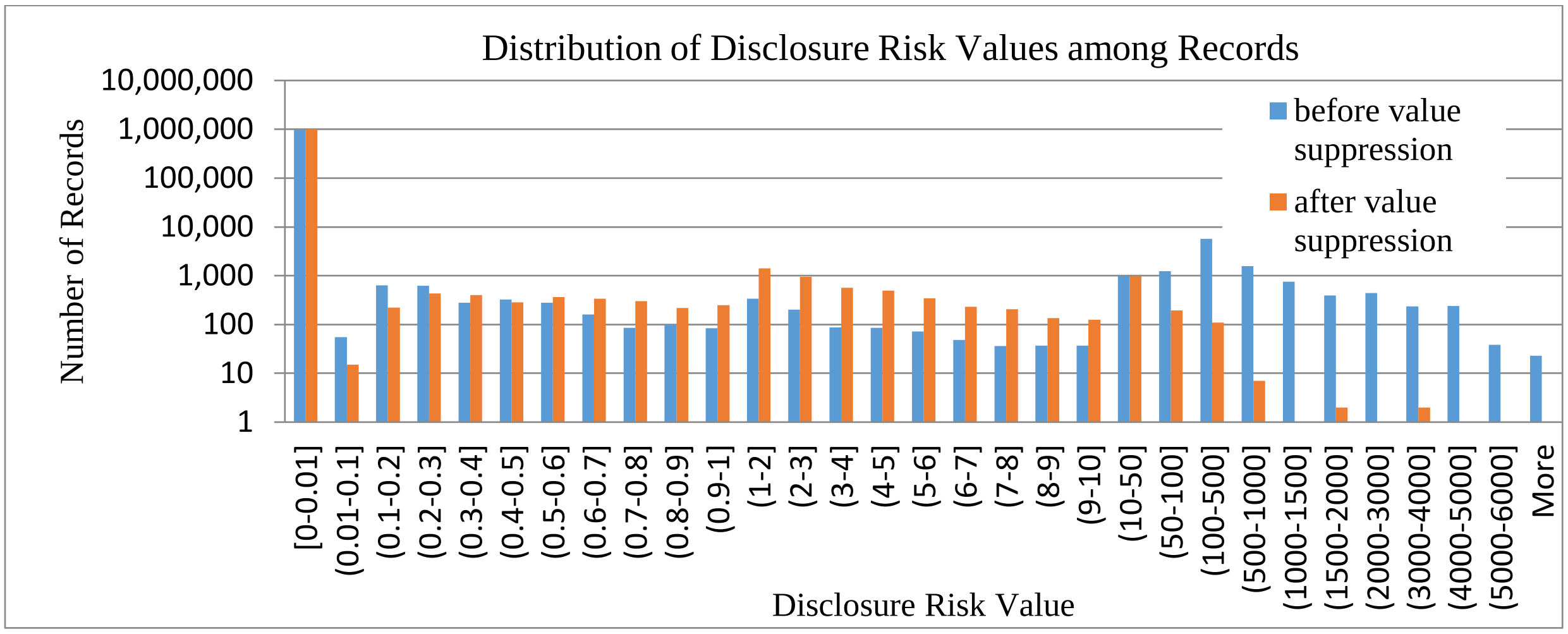

Figure 2. Comparing Histogram of Records Disclosure Risk Values before and after Value Suppression

**4.3 Data Utility Evaluation on the Anonymized dataset**

Since this dataset contains large number of SAs, this experiment implies the effectiveness of our approach on large number of SAs, in terms of preserving data utility. We have compared our obtained anonymized dataset with the anonymized dataset obtained from "k-anonymity", "$\ell$-Diversity", and "t-closeness" privacy models based on NCP measure. These privacy models are employed through ARX anonymization tool [14]. Looking at the assigned publicly known probabilities to each attribute, we have chosen the attributes with the probability of more than 0.01 as QIDs, which are 10 attributes. For "$\ell$-Diversity", and "t-closeness", which require specifying SAs, attributes other than QIDs are counted as SAs (17 number of SAs).

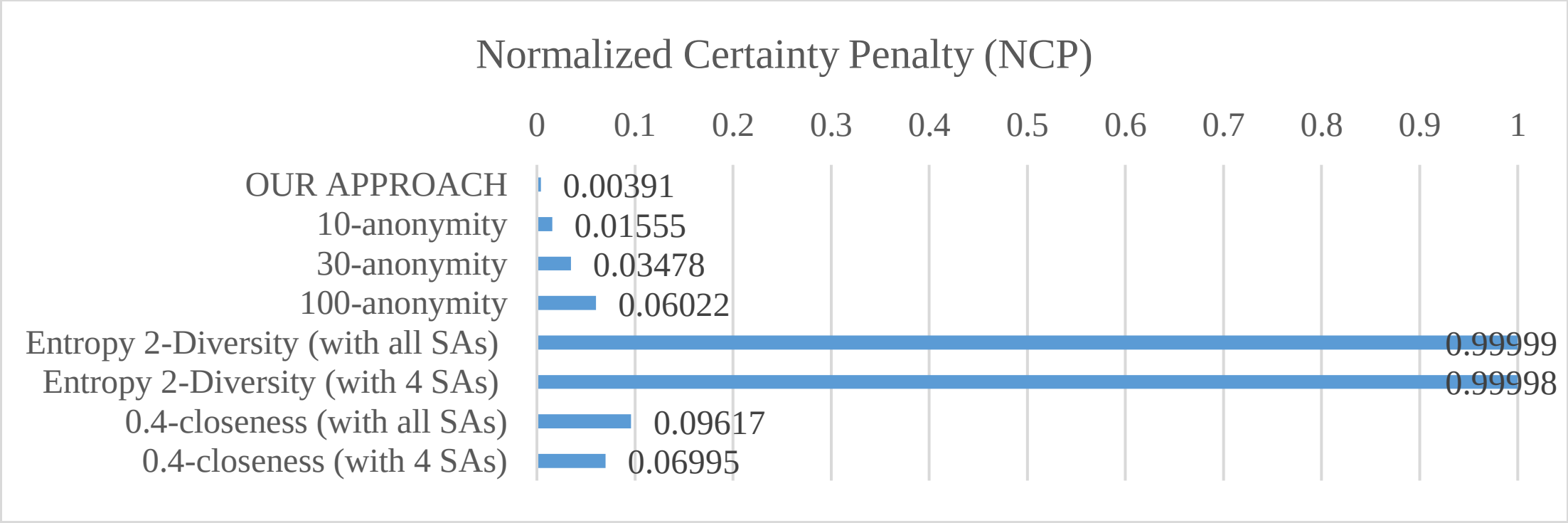

Figure 3. Comparing data utility of the anonymized dataset obtained from multiple approaches

Figure 3 indicates that the anonymized dataset obtained from our approach outperforms other approaches by having the least information loss. It is also shown that increasing k value in "k-anonymity" incurs more information loss, since it implies stricter privacy model. The reason why "ℓ-Diversity" incurs huge information loss is that the values of SAs in our dataset are highly skewed and it is already stated that ℓ-Diversity is not an effective model in these cases [3]. Moreover, this comparison denotes the fact that having more SAs leads to higher information loss.

## 5. Conclusions

In this paper, we have proposed a novel approach to reduce disclosure risk while preserving data utility. We presented a novel suppression method, which targets records at high risk of disclosure. We initially proposed a disclosure risk metric, which measures the risk of each record. Then value suppression algorithm is designed to suppress only the values of attributes, which have highest contribution in risk measure. By deleting these values, we are reducing the risk of high-risk records and since our proposed suppression algorithm considers only high-risk records, the other records staying intact and data utility is preserved. We have conducted an experiment on a confidential dataset with 19 SAs. Comparing the disclosure risk after suppression, the number of high-risk records is reduced by 45%. Besides, the information loss on the anonymized dataset is 0.39%, which indicates that our anonymization approach does not incur huge information loss on datasets with large number of SAs. Our proposed approach is dependent on a set of pre-defined parameters such as publicly known probabilities, sensitivity weights, and delta. In the future work, these parameters need to be characterized in terms of their effect on the disclosure risk values. Our approach can be extended for improving "Generalization", which is expected to enhance data utility, since generalization incurs less information loss compared to suppression.

## Acknowledgements

The authors would like to acknowledge the Social Research and Evaluation Center at Louisiana State University for providing this research a real-world sensitive dataset in social work domain.